\documentclass{Interspeech}
\usepackage{times}
\usepackage{latexsym}
\usepackage{booktabs}
\usepackage{verbatim}
\usepackage[table,xcdraw]{xcolor}
\usepackage{multirow}
\usepackage[table,xcdraw]{xcolor}
\usepackage{colortbl}  
\usepackage[table,xcdraw]{xcolor}
\usepackage{booktabs}
\usepackage[T1]{fontenc}
\usepackage[utf8]{inputenc}
\usepackage{pdfpages}
\usepackage{microtype}
\definecolor{lightred}{rgb}{1.0, 0.8, 0.8}
\definecolor{lightblue}{rgb}{0.8, 0.9, 1.0}
\definecolor{lightgreen}{rgb}{0.8, 1.0, 0.8}
\definecolor{lightyellow}{rgb}{1.0, 1.0, 0.8}
\definecolor{lightpurple}{rgb}{0.9, 0.8, 1.0}
\definecolor{lightorange}{rgb}{1.0, 0.9, 0.8}
\usepackage{inconsolata}
\usepackage{adjustbox}

\usepackage{etoolbox}
\makeatletter
\patchcmd{\affiliation}
  {\renewcommand{\affiliationsep}{\vskip1pt}}
  {\renewcommand{\affiliationsep}{, }}
  {}{}
\makeatother

\usepackage{ subcaption, algorithm2e, url, float, etoolbox, adjustbox, pgf,booktabs, verbatim}
\usepackage{xcolor}
\usepackage{amsmath}
\usepackage{amssymb}
\usepackage{tcolorbox}

\definecolor{mattemaroon5}{RGB}{255,245,250}   
\definecolor{mattemaroon10}{RGB}{250,230,235}  
\definecolor{mattemaroon15}{RGB}{245,215,220}  
\definecolor{mattemaroon20}{RGB}{240,200,205}  
\definecolor{mattemaroon25}{RGB}{235,185,190}  
\definecolor{mattemaroon30}{RGB}{230,170,175}  
\definecolor{mattemaroon35}{RGB}{225,155,160}  
\definecolor{mattemaroon40}{RGB}{220,140,145}  

\definecolor{matteGreen40}{RGB}{100,255,100} 
\definecolor{matteGreen35}{RGB}{120,255,120}
\definecolor{matteGreen30}{RGB}{140,255,140}
\definecolor{matteGreen25}{RGB}{160,255,160}
\definecolor{matteGreen20}{RGB}{180,255,180}
\definecolor{matteGreen15}{RGB}{210,255,210}
\definecolor{matteGreen10}{RGB}{240,255,240} 

\title{Enhancing In-Domain and Out-Domain EmoFake Detection via Cooperative Multilingual Speech Foundation Models}

\author[affiliation={1}]{Orchid Chetia}{Phukan*}
\author[affiliation={1,2}]{Mohd Mujtaba}{Akhtar*}
\author[affiliation={1,3}]{Girish*}{}
\author[affiliation={1}]{Arun Balaji}{Buduru}

\affiliation{}{IIIT-Delhi}{India}
\affiliation{}{V.B.S.P.U}{India}
\affiliation{}{UPES}{India}
\email{\textcolor{blue}{\texttt{Correspondence:}} orchidp@iiitd.ac.in} 

\usepackage{comment}

\begin{document}

\maketitle



\begin{abstract}

\noindent In this work, we address EmoFake Detection (EFD). We hypothesize that multilingual speech foundation models (SFMs) will be particularly effective for EFD due to their pre-training across diverse languages, enabling a nuanced understanding of variations in pitch, tone, and intensity. To validate this, we conduct a comprehensive comparative analysis of state-of-the-art (SOTA) SFMs. Our results shows the superiority of multilingual SFMs for same language (in-domain) as well as cross-lingual (out-domain) evaluation. To our end, we also propose, \textbf{\texttt{THAMA}} for fusion of foundation models (FMs) motivated by related research where combining FMs have shown improved performance. \textbf{\texttt{THAMA}} leverages the complementary conjunction of tucker decomposition and hadamard product for effective fusion. With \textbf{\texttt{THAMA}}, synergized with cooperative multilingual SFMs achieves topmost performance across in-domain and out-domain settings, outperforming individual FMs, baseline fusion techniques, and prior SOTA methods.

\end{abstract}

\section{Introduction}
\begin{figure}[!bt]
    \centering
    \includegraphics[width=1\linewidth]{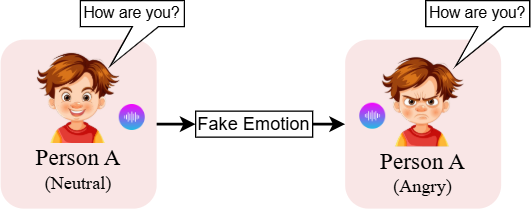}
    \caption{Demonstration of EmoFake: Person A's neutral speech is manipulated to synthesize an angry emotional tone in the audio while maintaining the same spoken content ('How are you?')}
    \label{fig:archi}
\end{figure}

Driven by the development of comprehensive datasets and rigorous benchmarking challenges, the field of fake audio detection has seen significant development. Early efforts, ASVspoof \cite{wu15e_interspeech}, laid the foundation for detecting spoofed speech, with subsequent iterations like ASVspoof 2017 \cite{kinnunen17_interspeech} and ASVspoof 2019 \cite{todisco19_interspeech} addressing replay attacks and synthetic speech generated via text-to-speech (TTS) and voice conversion (VC) systems. ASVspoof 2021 \cite{yamagishi21_asvspoof} further expanded its scope to encompass further more deepfake speech samples, reflecting the growing complexity of audio manipulation techniques. More recently, ADD challenges \cite{9746939, yi2023add} introduced datasets incorporating partially manipulated audio, fostering advancements in identifying hybrid fake audio scenarios. Further, Muller et al. \cite{muller2024mlaad} explores multi-lingual TTS-generated audio across diverse languages. Additionally, specialized tasks, such as singing voice deepfake detection \cite{10448184}, scenefake detection \cite{yi2024scenefake} have emerged, while the MLAAD dataset \cite{muller2024mlaad} explores multi-lingual TTS-generated audio across diverse languages. However, despite these advancements, an underexplored yet critical dimension remains: scenarios where the emotional state of speech is artificially modified i.e. EmoFake (EF) (Figure \ref{fig:archi}). While such technology holds promise for applications in entertainment, personalized speech systems, and human-computer interaction, its misuse raises serious concerns. EF can distort the intent of communication, sow distrust in audio-based media, and enable malicious applications, such as emotional manipulation, misinformation campaigns, and fraudulent schemes. For instance, altering the emotional tone of recorded evidence in legal contexts could undermine judicial integrity, while fabricating emotionally persuasive messages could exploit listeners’ psychological vulnerabilities at scale. These adverse implications highlight the urgent need for advanced detection frameworks that can effectively identify emotion-manipulated audio and mitigate its risks across sensitive domains, including security, media integrity, and forensics. As such Zhao et al. \cite{zhao2024emofake} presented the first database for EmoFake detection (EFD) and carried out an initial benchmarking. \par

\noindent Building on this foundation, we extend the exploration of EFD and investigate speech foundation models (SFMs) for EFD as SFMs holds immense potential for improving EFD also shown in previous research for audio deepfake detection \cite{kawa23b_interspeech, 10715076, stourbe24_asvspoof}. To our end, \textit{we hypothesize that multilingual SFMs will be most effective for EFD in comparison to other types of SFMs, as their broad linguistic pre-training will enable them to capture subtle emotional alterations in speech characteristics such as pitch, tone, intensity.} For validating our hypothesis, we present a comparative study of state-of-the-art (SOTA) SFMs including multilingual, monolingual, speaker recognition and music foundation models (MFMs). We are the first study to explore foundation models (FMs) for EFD to best of our knowledge. We include MFMs in our analysis as we thought that MFMs due to their pre-training on diverse music data may provide them capability of extracting pitch, timbre, etc. characteristics effectively important for detecting emotions and might be helpful for EFD. Further motivated by related research in speech recognition \cite{arunkumar22b_interspeech}, speech emotion recognition \cite{wu2023investigation}, synthetic speech detection \cite{chetia-phukan-etal-2024-heterogeneity}, where combination of FMs have shown improved performance, we investigate the fusion of FMs for EFD. We propose a novel framework, \textbf{\texttt{THAMA}} (\texttt{\textbf{T}}ucker-\texttt{\textbf{HA}}da\texttt{\textbf{MA}}rd Conjunction), which utilizes the synergistic combination of Tucker decomposition and Hadamard product to enable effective fusion of the FMs. \textbf{\texttt{THAMA}} with the fusion of cooperative multilingual SFMs approach sets a new benchmark for same-language (in-domain) and cross-lingual (out-domain) EFD. \par  
\noindent \textbf{To summarize, the main contributions of this work are as follows:}
\begin{itemize}
    \item To the best of our knowledge, we conduct the first large-scale and comprehensive comparative study of various SOTA SFMs, encompassing multilingual, monolingual, speaker recognition and music FMs for EFD. Our experiments shows that multilingual SFMs outperforms other FMs in both in-domain and out-domain evaluations. Our work marks the first-ever out-domain assessment of SOTA FMs for EFD also the first study exploring out-domain EFD. 
    \item We introduce a novel framework, \textbf{\texttt{THAMA}} for the fusion of FMs. With \textbf{\texttt{THAMA}}, through synchronization of multilingual SFMs, we achieve the best performance in comparison to individual FMs, baseline fusion techniques and setting new SOTA for EFD compared to previous SOTA works.
\end{itemize}
\noindent We will open-source the models and codes after the double-blind review for future research to build upon our work.

\section{Foundation Models}
In this section, we present the SOTA FMs used in our study. 

\noindent\textbf{Speech Foundation Models (SFMs)}: For monolingual SFMs, we consider WavLM\footnote{\url{https://huggingface.co/microsoft/wavlm-base}} \cite{chen2022wavlm}, Unispeech-SAT\footnote{\url{https://huggingface.co/microsoft/unispeech-sat-base}} \cite{chen2022unispeech}, Wav2vec2\footnote{\url{https://huggingface.co/facebook/wav2vec2-base}} \cite{baevski2020wav2vec}, and HuBERT\footnote{\url{https://huggingface.co/facebook/hubert-base-ls960}} \cite{hsu2021hubert}. 
WavLM and Unispeech-SAT are SOTA SFMs in SUPERB benchmark. WavLM is trained to solve speech denoising together with masked modeling whereas Unispeech-SAT is trained in a speaker-aware multi-task learning format. Wav2vec2 solves a contrastive learning objective and HuBERT solves BERT-like masked prediction loss. We make use of base versions of WavLM, Unispeech-SAT, wav2vec2, HuBERT with 94.70M, 94.68M, 95.04M, 94.68M parameters and all of them are pre-trained on librispeech 960 hours english data. For multilingual SFMs, we leverage XLS-R\footnote{\url{https://huggingface.co/facebook/wav2vec2-xls-r-300m}} \cite{babu22_interspeech}, Whisper\footnote{\url{https://huggingface.co/openai/whisper-base}} \cite{radford2023robust}, and MMS\footnote{\url{https://huggingface.co/facebook/mms-1b}} \cite{pratap2024scaling}. XLS-R was pre-trained on 128 languages, Whisper on 96 and MMS extended it to above 1400 languages. XLS-R and MMS follows wav2vec2 architecture and Whisper is built as a vanilla transformer encoder-decorder architecture. Whisper is trained in weakly-supervised manner and we use the base version with 74M parameters. For XLS-R and MMS, we use 300M and 1B parameters versions respectively. Further, we also include x-vector\footnote{\url{https://huggingface.co/speechbrain/spkrec-xvect-voxceleb}} \cite{8461375}, a speaker recognition SFM to our experiments. We believe it can be useful as it has shown its effectiveness in related areas such as synthetic speech detection \cite{chetia-phukan-etal-2024-heterogeneity} and speech emotion recognition \cite{9054317}. x-vector is a time-delay neural network trained on Voxceleb1 + Voxceleb2 for speaker recognition.

\noindent\textbf{Music Foundation Models (MFMs)}: We use MERT MFMs series \cite{li2023mert}. It includes different model variants including MERT-v1-330M\footnote{\url{https://huggingface.co/m-a-p/MERT-v1-330M}}, MERT-v1-95M\footnote{\url{https://huggingface.co/m-a-p/MERT-v1-95M}}, MERT-v0-public\footnote{\url{https://huggingface.co/m-a-p/MERT-v0-public}}, and MERT-v0\footnote{\url{https://huggingface.co/m-a-p/MERT-v0}} which are SOTA MFMs for extracting musical representations, achieving top performance in tasks like genre classification and music emotion recognition, trained on extensive and varied musical datasets. We also include \noindent music2vec-v1\footnote{\url{https://huggingface.co/m-a-p/music2vec-v1}} \cite{Music2VecAS}, a self-supervised SFM that excels in capturing detailed musical attributes and is widely applied in numerous music-related tasks. MERT-v1-330M contains 330M parameters and all the other MFMs contains 95M parameters. \par

\noindent MFMs needs the input audio to be resampled at sampling rates: MERT-v1-330M and MERT-v1-95M operate at 24 kHz, whereas MERT-v0-public, MERT-v0, and music2vec-v1 use 16 kHz. Similarly, SFMs are resampled to 16 kHz. Representations are extracted from frozen FMs using average pooling applied to the final hidden layer. The representation dimensions are 768 for music2vec-v1, most MERT variants, WavLM, Unispeech-SAT, Wav2vec2, and HuBERT; 1024 for MERT-v1-330M; 1280 for XLS-R and MMS; and 512 for x-vector and Whisper. Whisper representations are obtained exclusively from its encoder, omitting the decoder. \newline

\section{Modeling}
In this section, we present the downstream modeling approaches with individual FMs followed by the proposed framework, \textbf{\texttt{THAMA}} for combining the FMs. We use Fully Connected Network (FCN) and CNN as downstream networks with individual FMs. CNN starts with three 1D convolutional layers with 64, 128, 256 filters with a kernel size of 3 and ReLU activation. Each 1D convolutional layer is followed by max pooling, with a pool size of 2. Flattening is done and we attach a FCN block with 128 and 64 neurons using ReLU activation. The model performs binary classification for classifying fake \textit{vs} non-fake. The FCN model also consists the same dense layers as the CNN model FCN block. 

\begin{figure}[hbt!]
    \centering
    \includegraphics[width=0.75\linewidth]{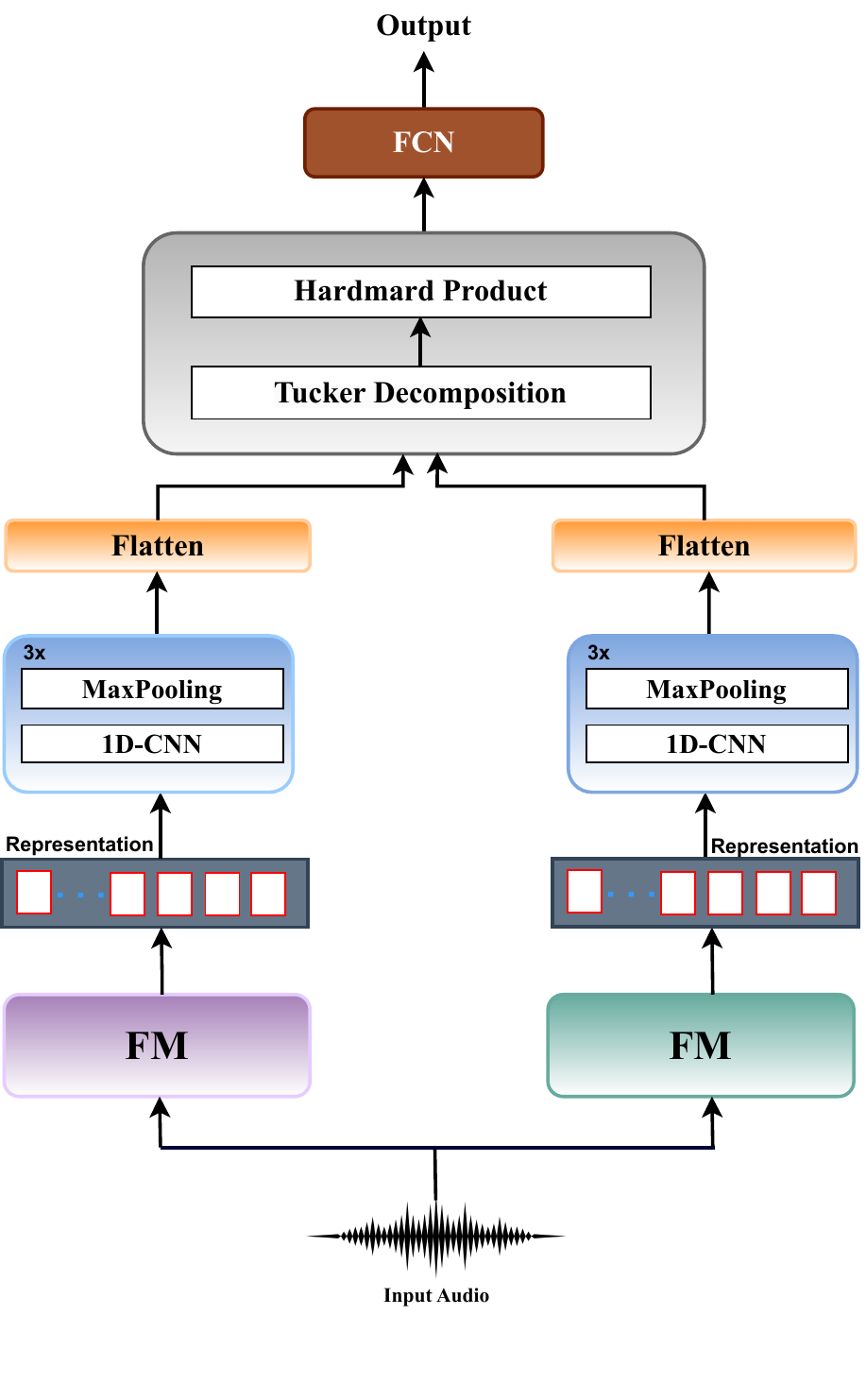}
    \caption{Proposed Framework: \textbf{\texttt{THAMA}}}
    \label{fig:proposed}
\end{figure}


\subsection{THAMA: Tucker-Hadamard Conjunction}

We present \textbf{THAMA} for fusion of FMs, with the modeling architecture shown in Figure \ref{fig:proposed}. Let \( \mathbf{X}_1 \in \mathbb{R}^{d_1} \) and \( \mathbf{X}_2 \in \mathbb{R}^{d_2} \) represent the representations from two FMs. These representations are first passed through a convolutional block with same architectural details (1D convolutional layers and maxpooling) as used with individual FMs above. The output of the convolutional block is then flattened, yielding \( \mathbf{X}_1' \in \mathbb{R}^{d_1'} \) and \( \mathbf{X}_2' \in \mathbb{R}^{d_2'} \). Next, these transformed representations are projected into a shared space as \( \mathbf{F}_1 = \mathbf{W}_1 \mathbf{X}_1' \) and \( \mathbf{F}_2 = \mathbf{W}_2 \mathbf{X}_2' \), where \( \mathbf{W}_1 \in \mathbb{R}^{d_1' \times d_f} \) and \( \mathbf{W}_2 \in \mathbb{R}^{d_2' \times d_f} \) are trainable weight matrices, and \( d_f \) is the dimensionality of the fused space. The projected features are then fused using \textbf{Tucker Decomposition}, where a core tensor \( \mathbf{T} \in \mathbb{R}^{d_f \times d_f \times d_f} \) models the high-order interactions between features. This fusion is computed as \( \mathbf{Z} = \sum_{i=1}^{d_f} \sum_{j=1}^{d_f} \sum_{k=1}^{d_f} T_{ijk} \cdot F_{1i} \cdot F_{2j} \), which can be written compactly as \( \mathbf{Z} = \mathbf{F}_1^\top \mathbf{T} \mathbf{F}_2 \). The Tucker Decomposition is particularly effective in fusion as it decomposes the core tensor into factorized matrices, reducing redundancy and enhancing representational capacity while maintaining computational efficiency. This approach captures fine-grained interactions across the two FMs representational space without excessive parameter overhead. The fused feature \( \mathbf{Z} \) undergoes an element-wise \textbf{Hadamard product}, resulting in \( \mathbf{H} = \mathbf{Z} \odot \mathbf{Z} \), where \( \odot \) denotes element-wise multiplication. The Hadamard product further enhances fusion by preserving element-wise interactions within the fused representation, strengthening feature dependencies and allowing the model to better capture complementary and discriminative information. \textbf{Tucker Decomposition with Hadamard Product} provides a novel and effective approach that leverages high-order fusion while preserving feature-level granularity, allowing the model to focus on both global and local information. Finally, the fused representation \( \mathbf{H} \) is passed through a FCN block with the same modeling details with the individual FMs above and finally followed by the output layer for binary classification. The number of trainable parameters in \texttt{\textbf{THAMA}} varies between 5.5M and 10M, depending on the dimension size of the input representation.

\begin{table}[hbt!]
\scriptsize
\centering
\setlength{\tabcolsep}{4pt} 
\begin{tabular}{l|c|c|c|c|c|c|c|c}
\toprule
            & \multicolumn{2}{c|}{\textbf{E}} & \multicolumn{2}{c|}{\textbf{C}} & \multicolumn{2}{c|}{\textbf{E(TR)-C(TE)}} & \multicolumn{2}{c}{\textbf{C(TR)-E(TE)}} \\
\cmidrule(lr){2-3} \cmidrule(lr){4-5} \cmidrule(lr){6-7} \cmidrule(lr){8-9}
\textbf{SFM} & \textbf{FCN} & \textbf{CNN} & \textbf{FCN} & \textbf{CNN} & \textbf{FCN} & \textbf{CNN} & \textbf{FCN} & \textbf{CNN} \\
\midrule
U           & \cellcolor{mattemaroon40}2.68 & \cellcolor{mattemaroon40}2.62 & \cellcolor{mattemaroon20}8.23 & \cellcolor{mattemaroon20}8.03 & \cellcolor{mattemaroon15}10.00 & \cellcolor{mattemaroon15}9.84 & \cellcolor{mattemaroon25}7.28 & \cellcolor{mattemaroon30}6.36 \\
W2          & \cellcolor{mattemaroon40}2.60 & \cellcolor{mattemaroon40}2.46 & \cellcolor{mattemaroon30}4.42 & \cellcolor{mattemaroon35}3.63 & \cellcolor{mattemaroon30}6.03 & \cellcolor{mattemaroon30}5.98 & \cellcolor{mattemaroon30}6.91 & \cellcolor{mattemaroon30}6.69 \\
W           & \cellcolor{mattemaroon30}5.88 & \cellcolor{mattemaroon35}5.14 & \cellcolor{mattemaroon20}8.82 & \cellcolor{mattemaroon35}5.06 & \cellcolor{mattemaroon15}13.71 & \cellcolor{mattemaroon20}9.61 & \cellcolor{mattemaroon10}14.86 & \cellcolor{mattemaroon10}13.34 \\
Hu          & \cellcolor{mattemaroon35}3.40 & \cellcolor{mattemaroon40}2.48 & \cellcolor{mattemaroon10}15.44 & \cellcolor{mattemaroon10}13.97 & \cellcolor{mattemaroon15}10.40 & \cellcolor{mattemaroon15}9.95 & \cellcolor{mattemaroon15}12.28 & \cellcolor{mattemaroon15}12.24 \\
\textbf{Wh}          & \cellcolor{mattemaroon40}\textbf{1.90} & \cellcolor{mattemaroon40}\textbf{1.89} & \cellcolor{mattemaroon35}\textbf{3.09} & \cellcolor{mattemaroon35}\textbf{2.68} & \cellcolor{mattemaroon35}\textbf{4.85} & \cellcolor{mattemaroon35}\textbf{3.42} & \cellcolor{mattemaroon30}\textbf{5.86} & \cellcolor{mattemaroon30}\textbf{5.63} \\
\textbf{X}          & \cellcolor{mattemaroon40}\textbf{1.84} & \cellcolor{mattemaroon40}\textbf{1.80} & \cellcolor{mattemaroon40}\textbf{2.14} & \cellcolor{mattemaroon40}\textbf{2.09} & \cellcolor{mattemaroon35}\textbf{4.03} & \cellcolor{mattemaroon35}\textbf{4.01} & \cellcolor{mattemaroon30}\textbf{5.06} & \cellcolor{mattemaroon30}\textbf{5.01}
\\
\textbf{MM}           & \cellcolor{mattemaroon40}\textbf{1.72} & \cellcolor{mattemaroon40}\textbf{1.69} & \cellcolor{mattemaroon40}\textbf{2.13} & \cellcolor{mattemaroon40}\textbf{2.04} & \cellcolor{mattemaroon35}\textbf{4.03} & \cellcolor{mattemaroon35}\textbf{3.39} & \cellcolor{mattemaroon30}\textbf{5.00} & \cellcolor{mattemaroon30}\textbf{4.95} \\
XV          & \cellcolor{mattemaroon35}5.14 & \cellcolor{mattemaroon35}5.11 & \cellcolor{mattemaroon35}5.46 & \cellcolor{mattemaroon35}5.45 & \cellcolor{mattemaroon25}7.48 & \cellcolor{mattemaroon30}6.46 & \cellcolor{mattemaroon25}7.44 & \cellcolor{mattemaroon25}7.35 \\
MT95        & \cellcolor{mattemaroon20}9.11 & \cellcolor{mattemaroon25}7.11 & \cellcolor{mattemaroon20}10.01 & \cellcolor{mattemaroon20}9.31 & \cellcolor{mattemaroon15}12.72 & \cellcolor{mattemaroon15}12.09 & \cellcolor{mattemaroon20}9.83 & \cellcolor{mattemaroon20}9.02  \\
MTP         & \cellcolor{mattemaroon15}10.95 & \cellcolor{mattemaroon20}8.86 & \cellcolor{mattemaroon10}17.85 & \cellcolor{mattemaroon10}15.05 & \cellcolor{mattemaroon10}18.08 & \cellcolor{mattemaroon10}16.25 & \cellcolor{mattemaroon15}14.26 & \cellcolor{mattemaroon15}11.80 \\
MT3M        & \cellcolor{mattemaroon25}6.46 & \cellcolor{mattemaroon35}3.57 & \cellcolor{mattemaroon20}9.68 & \cellcolor{mattemaroon30}6.60 & \cellcolor{mattemaroon20}9.84 & \cellcolor{mattemaroon20}9.81 & \cellcolor{mattemaroon20}9.40 & \cellcolor{mattemaroon25}7.60 \\
m2v         & \cellcolor{mattemaroon10}11.01 & \cellcolor{mattemaroon10}10.64 & \cellcolor{mattemaroon10}13.32 & \cellcolor{mattemaroon10}13.21 & \cellcolor{mattemaroon10}15.33 & \cellcolor{mattemaroon10}15.20 & \cellcolor{mattemaroon10}16.08 & \cellcolor{mattemaroon10}16.08 \\
MTV0        & \cellcolor{mattemaroon15}9.69 & \cellcolor{mattemaroon15}9.22 & \cellcolor{mattemaroon10}15.46 & \cellcolor{mattemaroon10}14.19 & \cellcolor{mattemaroon10}18.61 & \cellcolor{mattemaroon10}18.50 & \cellcolor{mattemaroon10}17.52 & \cellcolor{mattemaroon10}17.43 \\
\bottomrule
\end{tabular}
\caption{EER scores in \% for various FMs; E, C represents the english and chinese subset where the models were trained and tested on the same language; E(TR)-C(TE), C(TR)-E(TE) stands for training on english and testing on chinese, training on chinese and testing on english respectively; Abbreviations used for FMs: Unispeech-SAT - U, Wav2vec2 - W2, WavLM - W, Whisper - Wh, XLS-R - X, x-vector - XV, MMS - MM, HuBERT - Hu, music2vec-v1 - m2v, MERT-v1-95M - MT95, MERT-v0-public - MTP, MERT-v1-330M - MT3M, MERT-v0 - MTV0; The intensity highlights the performance levels, where darker shades indicate lower EER values and lighter shades indicate higher EER values; The abbreviations of the FMs and the intensity used in this table is kept same for Table \ref{fusion_table}}
\label{tab:eer_performance_single}
\end{table}

\begin{table}[hbt!]
\setlength{\tabcolsep}{12pt} 
\scriptsize
\centering
\adjustbox{max width=\textwidth}{
\begin{tabular}{lc|l|l|l}
\toprule
\multicolumn{1}{c|}{\multirow{3}{*}{}} & \multicolumn{2}{c|}{\textbf{Concat}} & \multicolumn{2}{c}{\textbf{\texttt{THAMA}}} \\
\cmidrule(lr){2-3} \cmidrule(lr){4-5}
\multicolumn{1}{c|}{} & \textbf{E} & \multicolumn{1}{c|}{\textbf{C}} & \multicolumn{1}{c|}{\textbf{E}} & \multicolumn{1}{c}{\textbf{C}} \\
\midrule
\multicolumn{1}{l|}{U + W2}    & \cellcolor{mattemaroon30}3.03 & \cellcolor{mattemaroon25}5.97  & \cellcolor{mattemaroon30}2.91 & \cellcolor{mattemaroon25}4.29 \\
\multicolumn{1}{l|}{U + W}     & \cellcolor{mattemaroon30}3.85 & \cellcolor{mattemaroon20}7.34  & \cellcolor{mattemaroon30}3.76 & \cellcolor{mattemaroon20}6.40 \\
\multicolumn{1}{l|}{U + Wh}    & \cellcolor{mattemaroon30}3.83 & \cellcolor{mattemaroon30}4.01  & \cellcolor{mattemaroon40}2.43 & \cellcolor{mattemaroon30}3.04 \\
\multicolumn{1}{l|}{U + X}     & \cellcolor{mattemaroon30}3.03 & \cellcolor{mattemaroon30}3.19  & \cellcolor{mattemaroon40}2.00 & \cellcolor{mattemaroon40}2.04 \\
\multicolumn{1}{l|}{U + XV}    & \cellcolor{mattemaroon30}3.89 & \cellcolor{mattemaroon25}4.83  & \cellcolor{mattemaroon35}2.58 & \cellcolor{mattemaroon35}2.45 \\
\multicolumn{1}{l|}{U + MM}    & \cellcolor{mattemaroon30}3.12 & \cellcolor{mattemaroon30}3.09  & \cellcolor{mattemaroon40}2.00 & \cellcolor{mattemaroon40}2.06 \\
\multicolumn{1}{l|}{U + Hu}    & \cellcolor{mattemaroon35}2.32 & \cellcolor{mattemaroon20}5.65  & \cellcolor{mattemaroon35}2.51 & \cellcolor{mattemaroon20}5.95 \\
\multicolumn{1}{l|}{U + m2v}   & \cellcolor{mattemaroon15}8.08 & \cellcolor{mattemaroon10}11.01  & \cellcolor{mattemaroon20}7.49 & \cellcolor{mattemaroon10}10.00 \\
\multicolumn{1}{l|}{U + MT95}  & \cellcolor{mattemaroon25}4.25 & \cellcolor{mattemaroon20}6.90  & \cellcolor{mattemaroon25}4.05 & \cellcolor{mattemaroon20}6.59 \\
\multicolumn{1}{l|}{U + MTP}   & \cellcolor{mattemaroon35}2.60 & \cellcolor{mattemaroon15}8.80  & \cellcolor{mattemaroon35}2.56 & \cellcolor{mattemaroon15}8.54 \\
\multicolumn{1}{l|}{U + MT3M}  & \cellcolor{mattemaroon35}2.35 & \cellcolor{mattemaroon20}7.22  & \cellcolor{mattemaroon25}4.20 & \cellcolor{mattemaroon20}7.30 \\
\multicolumn{1}{l|}{U + MTV0}  & \cellcolor{mattemaroon35}2.52 & \cellcolor{mattemaroon20}7.45  & \cellcolor{mattemaroon30}2.95 & \cellcolor{mattemaroon20}6.80 \\
\multicolumn{1}{l|}{W2 + W}    & \cellcolor{mattemaroon30}3.05 & \cellcolor{mattemaroon20}7.00  & \cellcolor{mattemaroon30}3.03 & \cellcolor{mattemaroon35}2.71 \\
\multicolumn{1}{l|}{W2 + Wh}   & \cellcolor{mattemaroon35}2.50 & \cellcolor{mattemaroon25}5.98  & \cellcolor{mattemaroon35}2.45 & \cellcolor{mattemaroon35}2.60 \\
\multicolumn{1}{l|}{W2 + X}    & \cellcolor{mattemaroon40}2.05 & \cellcolor{mattemaroon25}5.12  & \cellcolor{mattemaroon40}2.06 & \cellcolor{mattemaroon40}2.20 \\
\multicolumn{1}{l|}{W2 + XV}   & \cellcolor{mattemaroon40}2.22 & \cellcolor{mattemaroon25}5.94  & \cellcolor{mattemaroon40}2.17 & \cellcolor{mattemaroon30}3.26 \\
\multicolumn{1}{l|}{W2 + MM}    & \cellcolor{mattemaroon40}2.11 & \cellcolor{mattemaroon30}3.21  & \cellcolor{mattemaroon40}2.09 & \cellcolor{mattemaroon40}2.16 \\
\multicolumn{1}{l|}{W2 + Hu}   & \cellcolor{mattemaroon35}2.73 & \cellcolor{mattemaroon25}6.08  & \cellcolor{mattemaroon35}2.99 & \cellcolor{mattemaroon25}6.96 \\
\multicolumn{1}{l|}{W2 + m2v}  & \cellcolor{mattemaroon35}2.81 & \cellcolor{mattemaroon15}8.30  & \cellcolor{mattemaroon35}2.70 & \cellcolor{mattemaroon25}5.80 \\
\multicolumn{1}{l|}{W2 + MT95} & \cellcolor{mattemaroon30}3.51 & \cellcolor{mattemaroon20}6.85  & \cellcolor{mattemaroon25}4.32 & \cellcolor{mattemaroon25}4.58 \\
\multicolumn{1}{l|}{W2 + MTP}  & \cellcolor{mattemaroon30}3.85 & \cellcolor{mattemaroon20}6.08  & \cellcolor{mattemaroon30}3.83 & \cellcolor{mattemaroon30}3.38 \\
\multicolumn{1}{l|}{W2 + MT3M} & \cellcolor{mattemaroon25}4.30 & \cellcolor{mattemaroon10}9.91  & \cellcolor{mattemaroon25}5.15 & \cellcolor{mattemaroon30}3.43 \\
\multicolumn{1}{l|}{W2 + MTV0} & \cellcolor{mattemaroon25}4.30 & \cellcolor{mattemaroon10}9.71  & \cellcolor{mattemaroon25}4.23 & \cellcolor{mattemaroon30}3.50 \\
\multicolumn{1}{l|}{W + Wh}    & \cellcolor{mattemaroon30}3.57 & \cellcolor{mattemaroon30}4.08  & \cellcolor{mattemaroon30}3.61 & \cellcolor{mattemaroon30}3.05 \\
\multicolumn{1}{l|}{W + X}     & \cellcolor{mattemaroon35}2.32 & \cellcolor{mattemaroon30}4.03  & \cellcolor{mattemaroon40}2.00 & \cellcolor{mattemaroon40}2.03 \\
\multicolumn{1}{l|}{W + XV}    & \cellcolor{mattemaroon20}5.83 & \cellcolor{mattemaroon25}4.97  & \cellcolor{mattemaroon20}5.80 & \cellcolor{mattemaroon25}4.86 \\
\multicolumn{1}{l|}{W + MM}    & \cellcolor{mattemaroon40}2.00 & \cellcolor{mattemaroon30}4.03  & \cellcolor{mattemaroon40}2.00 & \cellcolor{mattemaroon40}2.03 \\
\multicolumn{1}{l|}{W + Hu}    & \cellcolor{mattemaroon35}2.91 & \cellcolor{mattemaroon10}14.34 & \cellcolor{mattemaroon40}2.06 & \cellcolor{mattemaroon10}13.57 \\
\multicolumn{1}{l|}{W + m2v}   & \cellcolor{mattemaroon15}8.08 & \cellcolor{mattemaroon10}12.00 & \cellcolor{mattemaroon15}8.09 & \cellcolor{mattemaroon10}12.06 \\
\multicolumn{1}{l|}{W + MT95}  & \cellcolor{mattemaroon25}4.06 & \cellcolor{mattemaroon10}16.31  & \cellcolor{mattemaroon25}4.14 & \cellcolor{mattemaroon20}6.99 \\
\multicolumn{1}{l|}{W + MTP}   & \cellcolor{mattemaroon20}6.25 & \cellcolor{mattemaroon15}11.06 & \cellcolor{mattemaroon20}6.28 & \cellcolor{mattemaroon15}11.19 \\
\multicolumn{1}{l|}{W + MT3M}  & \cellcolor{mattemaroon30}3.81 & \cellcolor{mattemaroon15}9.25  & \cellcolor{mattemaroon30}3.79 & \cellcolor{mattemaroon15}9.15 \\
\multicolumn{1}{l|}{W + MTV0}  & \cellcolor{mattemaroon25}4.86 & \cellcolor{mattemaroon15}12.16  & \cellcolor{mattemaroon25}5.09 & \cellcolor{mattemaroon15}11.68\\
\multicolumn{1}{l|}{\textbf{Wh + X}}  & \cellcolor{mattemaroon40}\textbf{1.67} & \cellcolor{mattemaroon40}\textbf{1.83} & \cellcolor{mattemaroon40}\textbf{1.53} & \cellcolor{mattemaroon40}\textbf{1.65} \\
\multicolumn{1}{l|}{Wh + XV}   & \cellcolor{mattemaroon30}3.11 & \cellcolor{mattemaroon35}2.89  & \cellcolor{mattemaroon40}2.03 & \cellcolor{mattemaroon40}2.02 \\
\multicolumn{1}{l|}{\textbf{Wh + MM}}  & \cellcolor{mattemaroon40}\textbf{1.38} & \cellcolor{mattemaroon40}\textbf{1.75} & \cellcolor{mattemaroon40}\textbf{1.24} & \cellcolor{mattemaroon40}\textbf{1.46} \\
\multicolumn{1}{l|}{Wh + Hu}   & \cellcolor{mattemaroon35}2.34 & \cellcolor{mattemaroon10}13.77 & \cellcolor{mattemaroon35}2.13 & \cellcolor{mattemaroon25}4.02 \\
\multicolumn{1}{l|}{Wh + m2v}  & \cellcolor{mattemaroon20}6.00 & \cellcolor{mattemaroon15}10.00 & \cellcolor{mattemaroon20}6.49 & \cellcolor{mattemaroon15}11.28 \\
\multicolumn{1}{l|}{Wh + MT95} & \cellcolor{mattemaroon30}3.40 & \cellcolor{mattemaroon10}13.34  & \cellcolor{mattemaroon30}3.27 & \cellcolor{mattemaroon20}5.82 \\
\multicolumn{1}{l|}{Wh + MTP}  & \cellcolor{mattemaroon40}2.14 & \cellcolor{mattemaroon15}8.55  & \cellcolor{mattemaroon35}2.43 & \cellcolor{mattemaroon30}3.18 \\
\multicolumn{1}{l|}{Wh + MT3M} & \cellcolor{mattemaroon30}3.35 & \cellcolor{mattemaroon15}8.51  & \cellcolor{mattemaroon25}4.38 & \cellcolor{mattemaroon25}4.86 \\
\multicolumn{1}{l|}{Wh + MTV0} & \cellcolor{mattemaroon30}3.75 & \cellcolor{mattemaroon15}9.93  & \cellcolor{mattemaroon30}3.95 & \cellcolor{mattemaroon25}4.08 \\
\multicolumn{1}{l|}{X + XV}    & \cellcolor{mattemaroon40}2.14 & \cellcolor{mattemaroon15}8.06  & \cellcolor{mattemaroon40}2.04 & \cellcolor{mattemaroon40}2.09 \\
\multicolumn{1}{l|}{\textbf{X + MM}}     & \cellcolor{mattemaroon40}\textbf{1.00} & \cellcolor{mattemaroon40}\textbf{1.17}  & \cellcolor{mattemaroon40}\textbf{0.89} & \cellcolor{mattemaroon40}\textbf{1.03} \\
\multicolumn{1}{l|}{X + Hu}    & \cellcolor{mattemaroon35}2.89 & \cellcolor{mattemaroon15}8.42  & \cellcolor{mattemaroon35}2.58 & \cellcolor{mattemaroon35}2.67 \\
\multicolumn{1}{l|}{X + m2v}   & \cellcolor{mattemaroon15}7.40 & \cellcolor{mattemaroon20}6.06  & \cellcolor{mattemaroon40}2.20 & \cellcolor{mattemaroon30}3.63 \\
\multicolumn{1}{l|}{X + MT95}  & \cellcolor{mattemaroon20}6.00 & \cellcolor{mattemaroon20}6.08  & \cellcolor{mattemaroon25}4.00 & \cellcolor{mattemaroon25}4.04 \\
\multicolumn{1}{l|}{X + MTP}   & \cellcolor{mattemaroon20}6.03 & \cellcolor{mattemaroon20}6.11  & \cellcolor{mattemaroon25}4.00 & \cellcolor{mattemaroon25}4.03 \\
\multicolumn{1}{l|}{X + MT3M}  & \cellcolor{mattemaroon20}7.01 & \cellcolor{mattemaroon20}6.82  & \cellcolor{mattemaroon25}4.03 & \cellcolor{mattemaroon25}4.04 \\
\multicolumn{1}{l|}{X + MTV0}  & \cellcolor{mattemaroon20}7.18 & \cellcolor{mattemaroon20}6.05  & \cellcolor{mattemaroon25}5.13 & \cellcolor{mattemaroon20}6.03 \\
\multicolumn{1}{l|}{MM + Hu}   & \cellcolor{mattemaroon30}3.75 & \cellcolor{mattemaroon15}9.82  & \cellcolor{mattemaroon40}2.03 & \cellcolor{mattemaroon40}2.03 \\
\multicolumn{1}{l|}{MM + m2v}  & \cellcolor{mattemaroon20}6.00 & \cellcolor{mattemaroon10}16.05 & \cellcolor{mattemaroon25}4.51 & \cellcolor{mattemaroon25}5.51 \\
\multicolumn{1}{l|}{MM + MT95} & \cellcolor{mattemaroon20}6.20 & \cellcolor{mattemaroon15}9.17  & \cellcolor{mattemaroon25}4.78 & \cellcolor{mattemaroon25}5.50 \\
\multicolumn{1}{l|}{MM + MTP}  & \cellcolor{mattemaroon20}6.03 & \cellcolor{mattemaroon15}9.06  & \cellcolor{mattemaroon20}5.95 & \cellcolor{mattemaroon15}9.01 \\
\multicolumn{1}{l|}{MM + MT3M} & \cellcolor{mattemaroon20}7.00 & \cellcolor{mattemaroon15}9.04  & \cellcolor{mattemaroon20}6.95 & \cellcolor{mattemaroon20}6.32 \\
\multicolumn{1}{l|}{MM + MTV0} & \cellcolor{mattemaroon20}6.89 & \cellcolor{mattemaroon15}9.05  & \cellcolor{mattemaroon20}7.34 & \cellcolor{mattemaroon20}7.72 \\
\multicolumn{1}{l|}{Hu + m2v}   & \cellcolor{mattemaroon10}10.59& \cellcolor{mattemaroon10}12.28 & \cellcolor{mattemaroon25}5.80 & \cellcolor{mattemaroon20}6.63 \\
\multicolumn{1}{l|}{Hu + MT95}  & \cellcolor{mattemaroon25}5.48 & \cellcolor{mattemaroon10}13.13 & \cellcolor{mattemaroon25}5.59 & \cellcolor{mattemaroon10}10.29 \\
\multicolumn{1}{l|}{Hu + MTP}   & \cellcolor{mattemaroon25}5.03 & \cellcolor{mattemaroon10}13.06 & \cellcolor{mattemaroon25}5.11 & \cellcolor{mattemaroon10}14.05 \\
\multicolumn{1}{l|}{Hu + MT3M}  & \cellcolor{mattemaroon25}5.77 & \cellcolor{mattemaroon10}13.68 & \cellcolor{mattemaroon25}5.00 & \cellcolor{mattemaroon10}10.03 \\
\multicolumn{1}{l|}{Hu + MTV0}  & \cellcolor{mattemaroon25}6.14 & \cellcolor{mattemaroon10}14.60 & \cellcolor{mattemaroon25}6.12 & \cellcolor{mattemaroon15}9.02 \\
\multicolumn{1}{l|}{m2v + MT95} & \cellcolor{mattemaroon10}14.01& \cellcolor{mattemaroon10}17.24 & \cellcolor{mattemaroon15}8.03 & \cellcolor{mattemaroon15}9.04 \\
\multicolumn{1}{l|}{m2v + MTP}  & \cellcolor{mattemaroon10}14.35& \cellcolor{mattemaroon10}15.65 & \cellcolor{mattemaroon15}8.03 & \cellcolor{mattemaroon15}9.11 \\
\multicolumn{1}{l|}{m2v + MT3M} & \cellcolor{mattemaroon10}13.71& \cellcolor{mattemaroon10}14.76 & \cellcolor{mattemaroon15}7.33 & \cellcolor{mattemaroon15}9.11 \\
\multicolumn{1}{l|}{m2v + MTV0} & \cellcolor{mattemaroon10}14.79& \cellcolor{mattemaroon15}10.00 & \cellcolor{mattemaroon15}8.57 & \cellcolor{mattemaroon15}9.09 \\
\multicolumn{1}{l|}{MT95 + MTP} & \cellcolor{mattemaroon20}7.79 & \cellcolor{mattemaroon15}9.80  & \cellcolor{mattemaroon20}7.63& \cellcolor{mattemaroon15}9.46 \\
\multicolumn{1}{l|}{MT95 + MT3M}& \cellcolor{mattemaroon20}7.08 & \cellcolor{mattemaroon15}8.02  & \cellcolor{mattemaroon20}6.85 & \cellcolor{mattemaroon10}12.02 \\
\multicolumn{1}{l|}{MT95 + MTV0}& \cellcolor{mattemaroon20}7.35 & \cellcolor{mattemaroon10}12.98 & \cellcolor{mattemaroon20}7.28 & \cellcolor{mattemaroon10}12.41 \\
\multicolumn{1}{l|}{MTP + MT3M} & \cellcolor{mattemaroon15}8.19 & \cellcolor{mattemaroon15}11.54 & \cellcolor{mattemaroon15}8.14 & \cellcolor{mattemaroon15}10.65 \\
\multicolumn{1}{l|}{MTP + MTV0} & \cellcolor{mattemaroon15}8.72 & \cellcolor{mattemaroon10}15.74 & \cellcolor{mattemaroon15}8.68 & \cellcolor{mattemaroon10}14.32 \\
\multicolumn{1}{l|}{MT3M + MTV0}& \cellcolor{mattemaroon20}7.40 & \cellcolor{mattemaroon10}12.83 & \cellcolor{mattemaroon20}7.35 & \cellcolor{mattemaroon15}11.66 \\
\bottomrule
\end{tabular}}
\caption{Evaluations scores for combinations of different FMs; EER scores are given in \%; E, C represents English and Chinese}
\label{fusion_table}
\end{table}

\section{Experiments}
\subsection{Benchmark Dataset}
We use the only dataset available for EFD by Zhao et al. \cite{zhao2024emofake}. It comprises recordings in both English and Chinese, featuring audio samples from multiple speakers across five core emotions: Neutral, Happy, Angry, Sad, and Surprise. Fake emotional samples are generated using seven open-source Emotional Voice Conversion models which alter emotional expressions while retaining speaker identity. The dataset is divided into Chinese and English subsets, each with distinct training, development, and test splits. The training, development, and test contains 27300, 9100, and 17500 samples for both English and Chinese. 




\noindent\textbf{Training Details}: We train the models using the Adam optimizer with an initial learning rate of 1e-3, leveraging a batch size of 32 and 100 epochs. We use binary cross-entropy as the loss function. We use dropout, reduce learning rate and early stopping to prevent overfitting. We train and evaluate the models on the official split given by \cite{zhao2024emofake}. 



\subsection{Experimental Results}
\begin{figure}[!ht]
    \centering
    \subfloat[]{%
        \includegraphics[width=0.23\textwidth]{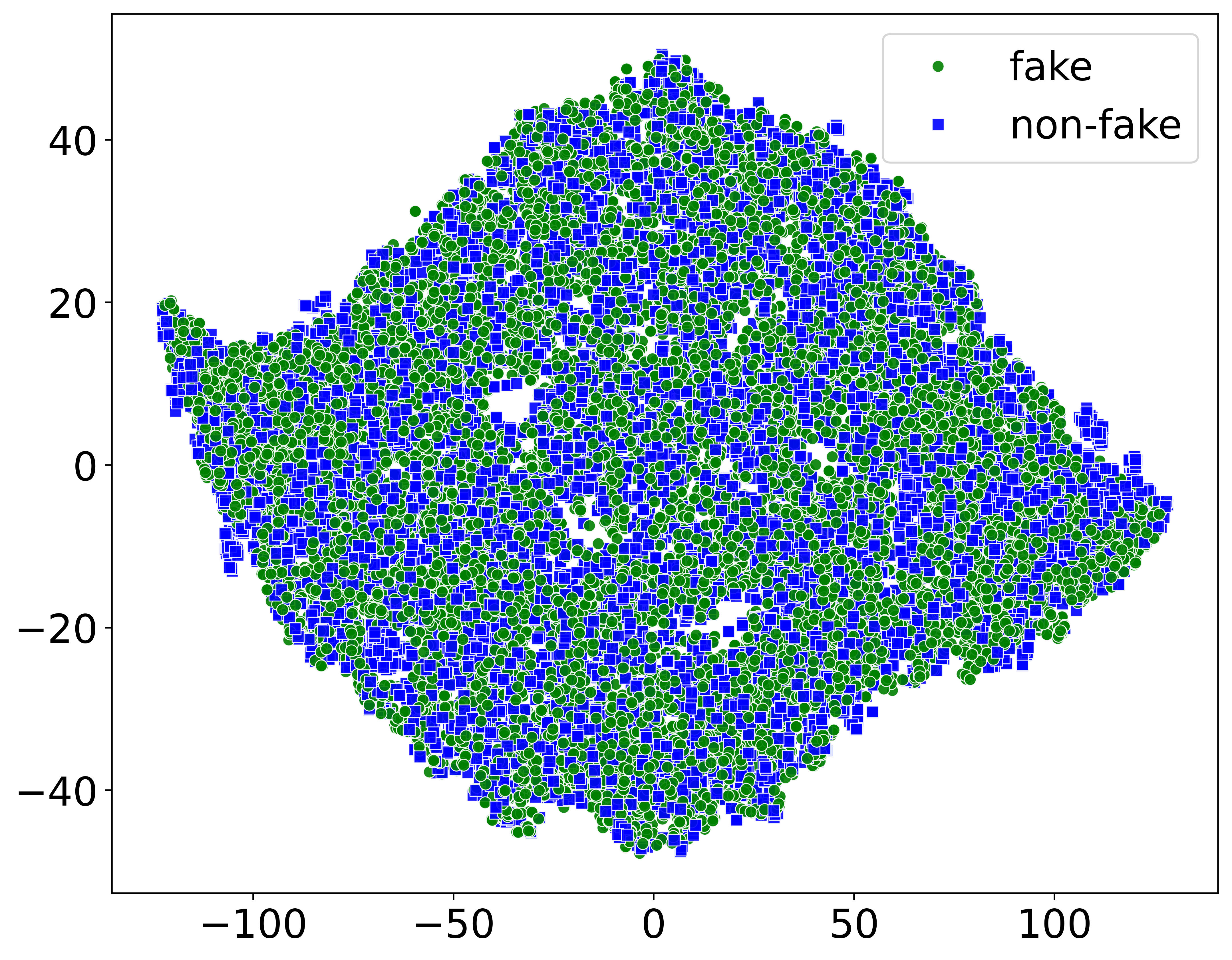}
    }
    \hfill
    \subfloat[]{%
        \includegraphics[width=0.23\textwidth]{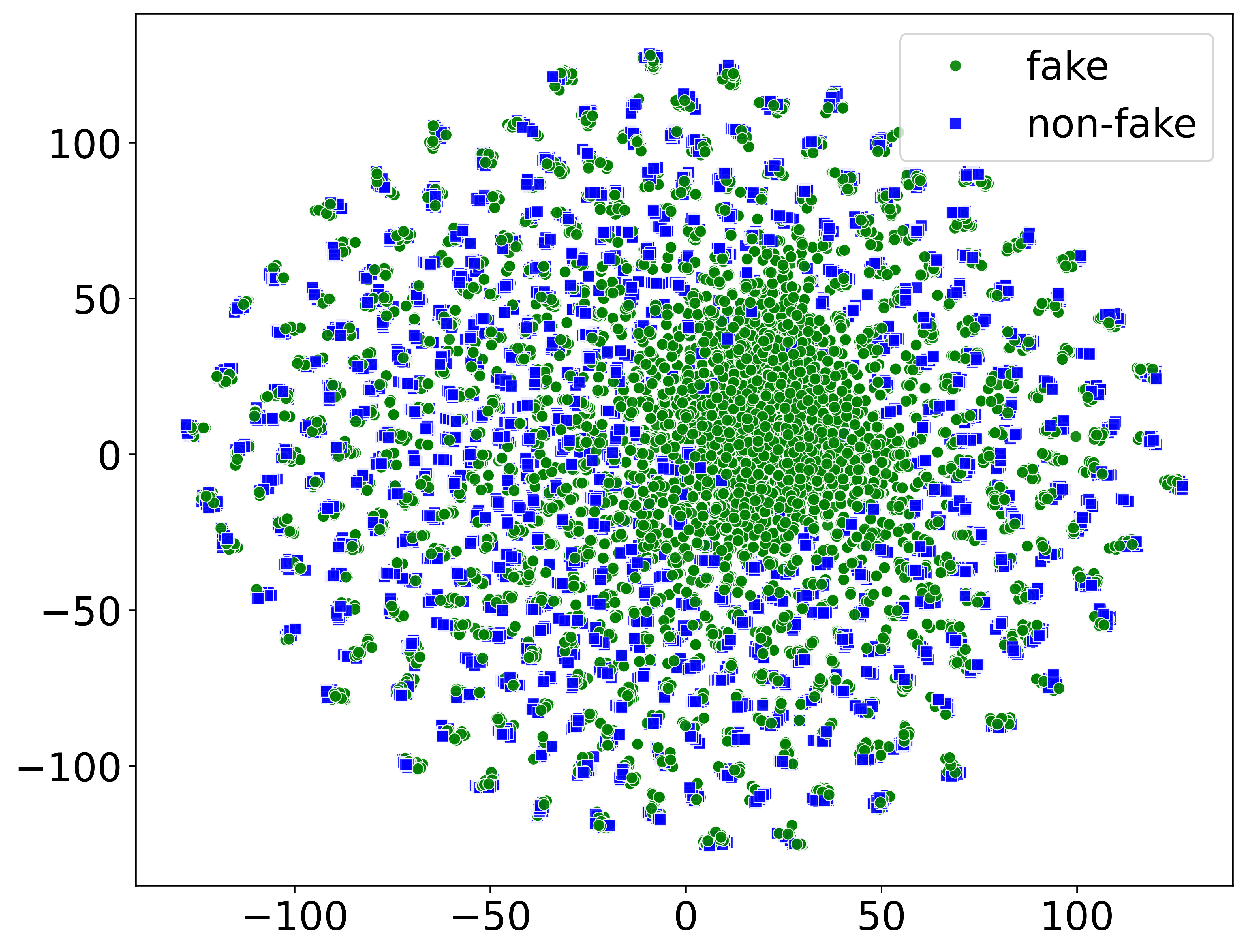}
    }
    \\
    \subfloat[]{%
        \includegraphics[width=0.23\textwidth]{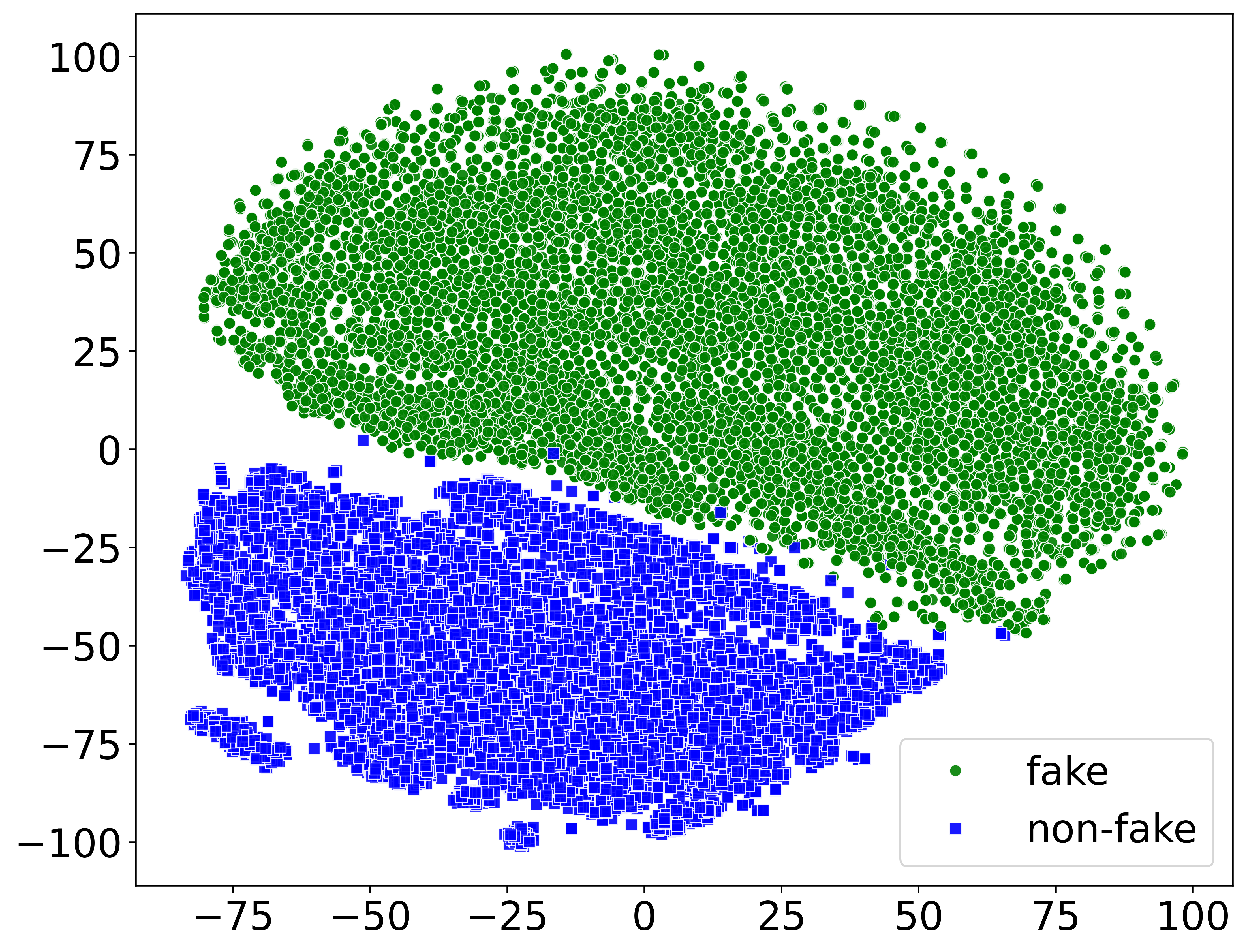}
    }
    \hfill
    \subfloat[]{%
        \includegraphics[width=0.23\textwidth]{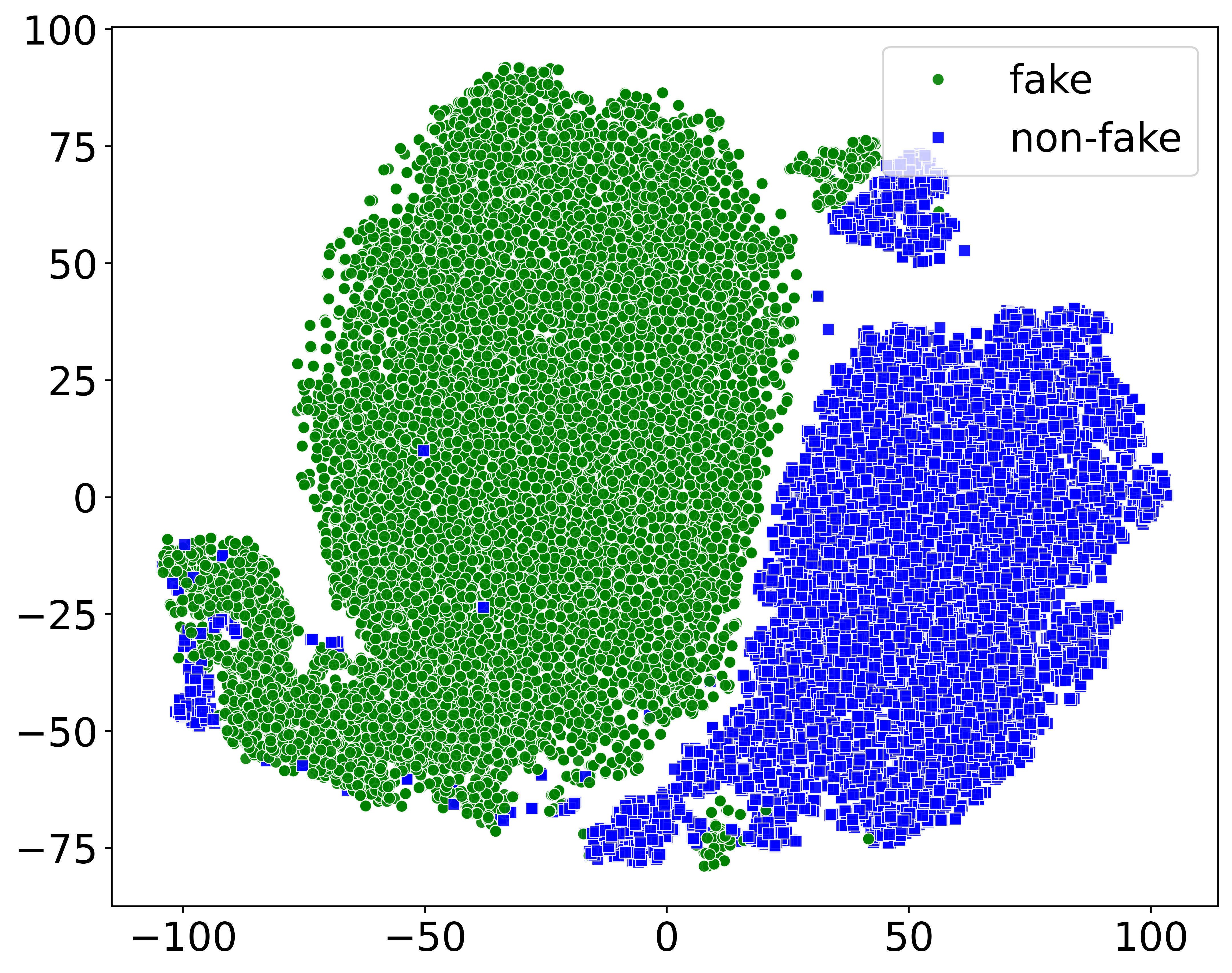}
    }
    \caption{t-SNE Plots- (a) MERT-v1-330M(b) WavLM (c) MMS (d) XLS-R}
    \label{fig:tsne}
\end{figure}

We use EER as the evaluation metric as used by previous resarch on EFD \cite{zhao2024emofake}. Table \ref{tab:eer_performance_single} presents the evaluation results of downstream models trained with individual FMs representations. For out-domain, we train on the training set of a particular language and test on the testing set of the other language. From the results, it is evident that multilingual SFMs consistently outperform monolingual, speaker recognition SFMs as well as MFMs in both in-domain and out-domain scenarios. This performance validates \textit{our hypothesis that multilingual SFMs will be particularly effective for EFD, as their diverse linguistic pre-training enables them for more nuanced understanding of variations in emotional cues through variations in speech characteristics such pitch, tone, and intensity.} Among the multilingual SFMs, MMS reported the topmost performance and can be attributed to its larger model size that provides them with much better understanding of emotional alterations. Within the monolingual SFMs, Wav2vec2 reports relatively well in comparison to its other monolingual counterparts and HuBERT performed the worst. The MFMs reported lower performance in comparison to other SFMs with music2vec-v1 reporting the worst among them all. Thus, showing its ineffectiveness in capturing emotional variations for better EFD. Overall, the CNN downstreams performed better than FCN models. We also plot the t-SNE visualization of the raw FMs representations in Figure \ref{fig:tsne}. We observe better clustering across the classes for multilingual SFMs, thus supporting our obtained results. Table \ref{fusion_table} shows the evaluation scores with fusion of different FMs. We show that experiments with \textbf{\texttt{THAMA}} leads to much improved performance than baseline concatenation-based fusion tehcnique. For baseline modeling, we follow the same modeling architecture till flattening as in Figure \ref{fig:archi}, after flattening, we use concatenation as fusion mechanism followed by FCN block as used with \textbf{\texttt{THAMA}}. Fusion of multilingual SFMs with \texttt{\textbf{THAMA}} reported the topmost performance compared to various combinations of FMs. This shows the emergence of complementary behavior amongst the multilingual SFMs and thus improving EFD in both English and Chinese. The results shown in Table \ref{fusion_table} are in-domain evaluations. We evaluate our best performing pair of multilingual SFMs (XLS-R + MMS) with \textbf{\texttt{THAMA}} and baseline concatenation-based fusion for out-domain scenario. We report an EER of 3.01\% when trained on English and tested on Chinese for \textbf{\texttt{THAMA}}. For training on Chinese and testing on English, \textbf{\texttt{THAMA}} reported EER of 4.21\%. For baseline concatenation-based fusion with XLS-R + MMS, we got EER of 3.33\% (training on english and testing on chinese) and 4.86 \%(training on chinese and testing on english). Results with \textbf{\texttt{THAMA}} outperforms both individual FMs and baseline fusion technique This shows the superiority of \textbf{\texttt{THAMA}} for effective fusion of FMs for in-domain and out-domain EFD.

\noindent \textbf{Comparison to SOTA}: We compare our best performing model \textbf{\texttt{THAMA}} with fusion of XLS-R and MMS with previous SOTA work on EFD \cite{zhao2024emofake}. They reported their best scores for in-domain evaluations as 3.65 \% and 8.34 \% EER on English and Chinese. We report the best scores as 0.89 \% and 1.03 \% EER on English and Chinese. With these scores, we set the new SOTA for EFD. Due to space constraint we are unable to present the comparison with SOTA in tabular format. For out-domain, no work has explored out-domain EFD yet and our work sets a benchmark for future works to work towards this direction.

\vspace{-0.5cm}
\section{Conclusion}
In this work, we show the effectiveness of multilingual SFMs for EFD in comparison to other SOTA SFMs and MFMs and demonstrating their ability to capture subtle emotional variations through their diverse linguistic pre-training. Our comprehensive analysis confirms their superiority in both in-domain and cross-lingual settings. Also, we introduce \texttt{\textbf{THAMA}}, a novel fusion framework that integrates Tucker decomposition and the Hadamard product for fusion of FMs. By combining THAMA with XLS-R and MMS, we achieve SOTA performance, surpassing individual models, baseline fusion techniques, and prior approaches. These findings highlight the potential of leveraging multilingual SFMs and their complementary behavior for improved EFD.

\bibliographystyle{IEEEtran}
\bibliography{mybib}

\end{document}